\documentclass[lettersize,journal]{IEEEtran}
\usepackage{amsmath,amsfonts}

\usepackage[linesnumbered,ruled]{algorithm2e}

\usepackage{array}
\usepackage[caption=false,font=normalsize,labelfont=sf,textfont=sf]{subfig}
\usepackage{textcomp}
\usepackage{stfloats}
\usepackage{url}
\usepackage{verbatim}
\usepackage{graphicx}
\usepackage{cite}
\usepackage{multirow}
\usepackage{amssymb} 
\usepackage{cases} 
\usepackage{makecell} 
\usepackage{diagbox} 
\usepackage{hyperref}

\begin{document}
	
	\title{Channel Sounding Using Multiplicative Arrays Based on Successive Interference Cancellation Principle }
	{\author{Zhangzhang Jiang, Zhiqiang Yuan, Chunhui Li, Le Yu, and Wei Fan
		\thanks{Zhangzhang Jiang, Zhiqiang Yuan, Chunhui Li, Le Yu, and Wei Fan are with the National Mobile Communications Research Laboratory, School of Information Science and Engineering, Southeast University, Nanjing 210096, China (email: \{jiangzhangzhang,zqyuan,lichunhui,yule,weifan\}@seu.edu.cn). \textit{Corresponding author: Wei Fan.}}
		}
		
		
		
		\maketitle
		
		\begin{abstract}
			
			Ultra-massive multiple-input and multiple-output (MIMO) systems have been seen as the key radio technology for the advancement of wireless communication systems, due to its capability to better utilize the spatial dimension of the propagation channels. Channel sounding is essential for developing accurate and realistic channel models for the massive MIMO systems. However, channel sounding with large-scale antenna systems has faced significant challenges in practice. The real antenna array based (RAA) sounder suffers from high complexity and cost, while virtual antenna array (VAA) solutions are known for its long measurement time. Notably, these issues will become more pronounced as the antenna array configuration gets larger for future radio systems. In this paper, we propose the concept of  multiplicative array (MA) for channel sounding applications to achieve large antenna aperture size with reduced number of required antenna elements. The unique characteristics of the MA are exploited for wideband spatial channel sounding purposes, supported by both one-path and multi-path numerical simulations. To address the fake paths and distortion in the angle delay profile issues inherent for MA in multipath channel sounding, a novel channel parameter estimation algorithm for MA based on successive interference cancellation (SIC) principle is proposed. Both numerical simulations and experimental validation results are provided to demonstrate the effectiveness and robustness of the proposed SIC algorithm for the MA. This research contributes significantly to the channel sounding and characterization of massive MIMO systems for future applications.
		\end{abstract}
		
		
		\section{Introduction}
		\IEEEPARstart{W}{ith} the evolution of wireless communication technologies, antenna arrays have become increasingly large and sophisticated. Massive MIMO as an emerging technology significantly amplifies the capabilities of conventional MIMO systems, potentially increasing their scale by several orders of magnitude \cite{larsson2014massive}. This allows antenna arrays to more effectively exploit the spatial dimension and enhance spectral efficiency. As fifth-generation (5G) systems transition into commercial reality, there is already substantial interest in the development of systems beyond 5G, commonly referred to as the sixth generation (6G) of wireless communication systems \cite{tataria20216g} which will further exploit the advancements in MIMO technology, referred to as gigantic MIMO (also referred to as ultra-massive (MIMO). It is envisioned that arrays with over one thousand antennas will be accommodated for low and mid-frequency bands, and many more for high-frequency bands such as millimeter wave and terahertz (THz) frequency bands \cite{10298067}. 
		
		Accurate channel modeling is essential for the design and performance evaluation of massive MIMO systems. Acquiring channel data from deployment scenarios and meticulously extracting multipath components (MPCs) are foundational for constructing precise and realistic channel models. However, as the array size increasingly expands, it brings huge challenges for channel measurement using antenna arrays (whether real \cite{lyu2023sub}, virtual \cite{lyu2023virtual}, or multiplexed \cite{zeng2016millimeter}),  due to potential increases in cost, complexity, and measurement time associated with the employed channel sounders which are specifically designed to record channel responses within the deployment scenario. 
		
		Directional scan sounding (DSS) utilizes a highly directional antenna with a rotator to directly record signals from various angles, offering a simple and cost-effective sounding scheme \cite{samimi201328}. Multiple channel sounding campaigns are conducted at millimeter-wave (mm-wave) band using this scheme \cite{7954664}. A modular correlation-based sub-THz channel sounder based on DSS is proposed in \cite{zhang2024modular}. However, it suffers from limited spatial resolution and long measurement time \cite{10132976}. Various multi-antenna channel sounding schemes have been explored in recent studies. The real antenna array (RAA) based sounder, utilizing individual RF chains for each antenna with a digital beamforming structure, enables simultaneous real-time channel response capture across all antenna elements \cite{7063445}. Virtual antenna array (VAA) based scheme employs a mechanically positioned omni-directional or directional antenna moved to various spatial locations to record channel responses, offering a generic, scalable solution for all frequency bands \cite{10439208}. A solution proposed in \cite{9322491} used phased array antenna \cite{caudill2019omnidirectional} via pre-defined amplitude and phase assignments to each array element for rapid 360$^{\circ}$ coverage. Cai \textit{et al.} designed a dynamic
		millimeter-wave channel sounder based on switched antenna array (SAA) principle \cite{cai2024switched}.

		Despite their practicality for massive MIMO channel measurements in static environments, VAA-based sounders are limited by long measurement times \cite{10439208, medbo201560} , while SAA \cite{9895366, li2019massive}, phased array \cite{caudill2019omnidirectional} and the RAA-based sounders also face extremely high system cost and signal processing complexity with large-scale arrays. Consequently, there still remains a lack of channel sounders capable of ultra-massive MIMO measurements in the deployment scenarios. Moreover, implementing these sounders in MIMO over-the-air (OTA) testing introduces additional challenges. In the multi-probe anechoic chamber (MPAC) setups for MIMO OTA testing, validation of the emulated channels in the test zone is essential to esnure that the target propagation channels are accurately emulated within the test zone\cite{gao2022over}. Different virtual array schemes have been standardized, e.g.  virtual uniform linear array (ULA) for the Long Term Evolution (LTE) \cite{3gpp.37.977},  virtual non-uniform circular array for 5G new radio (NR) frequency range (FR) one and 3D virtual arrays for 5G NR FR2 \cite{3gpp.38.827}. Significant efforts have been taking in standardization to reduce the number of required virtual element locations to improve the measurement efficiency. As the antenna array configurations increase, the measurement time for VAA and the measurement cost for RAA solutions will become more problematic.
		
		For antenna synthesis application, the concept of multiplicative array (MA), which basically utilizes two conventionally orthogonal linear arrays to replace uniform rectangular array (URA), addresses the large scale array problems, as proposed in \cite{aumann2010pattern}. It utilizes significantly less antenna elements than a URA but can achieve the same array synthesis pattern as URA, which has been widely used in radio astronomy \cite{grill2003new, slattery1966use} due to its superior angular resolution and cost-effectiveness.

		
		The MA concept can be valuable to significantly reduce the multi-antenna resource while maintaining the array performance. However, to the best of our knowledge, there has been currently no reports of the application of the MA in channel sounding. Therefore, this work aims to fill the above gap. The utilization of the MA in channel sounding will effectively reduce the cost, complexity for the RAA-based scheme, and measurement time for the VAA-based scheme, thus effectively addressing the demanding requirement of excessive number of antenna elements in massive MIMO channel sounding systems.  
		
		The main contributions of this paper are listed as follows:
		\begin{itemize} 
			\item Firstly, we introduce the concept of the MA for channel sounding, and demonstrate its performance for both one-path and multi-path scenarios with theoretical derivations and supported simulation results. For the first time, we point out that the MA concept has the potential for channel sounding and it will introduce fake paths with ambiguous angular positions and delays in multi-path scenario.
			
			\item Subsequently, we propose a multipath estimation algorithm tailored for the MA channel sounding based on the successive interference cancellation (SIC) principle to solve the problems in the multi-path scenario.
			
			\item Finally, we conduct experimental campaigns to validate the MA concept and prove the effectiveness of the proposed multipath parameter estimation algorithm.
			
		\end{itemize}
		
		The rest of this article is organized as follow. In Section II, we revisit the pattern synthesis techniques for the MA, and demonstrate the application of the MA in channel sounding and evaluate the performance in both angular and delay domain. In Section III, the proposed channel parameter estimation algorithm based on SIC is introduced in detail with corresponding simulations. In Section IV, the indoor experimental validation measurement and results analysis are provided, while the findings are concluded in Section V.

		\section{Signal Model for Multiplicative Array}
		In this section, we first review pattern synthesis theory for the MA. Subsequently, we introduce the application of the MA into channel sounding and demonstrate the performance in the wideband multi-path scenario, providing a detailed derivation supported by simulation results. Finally, we highlight the origin of the fake paths for the MA in channel sounding. 
		
		\begin{figure}[!t]
			\centering
			\includegraphics[width=0.5\textwidth]{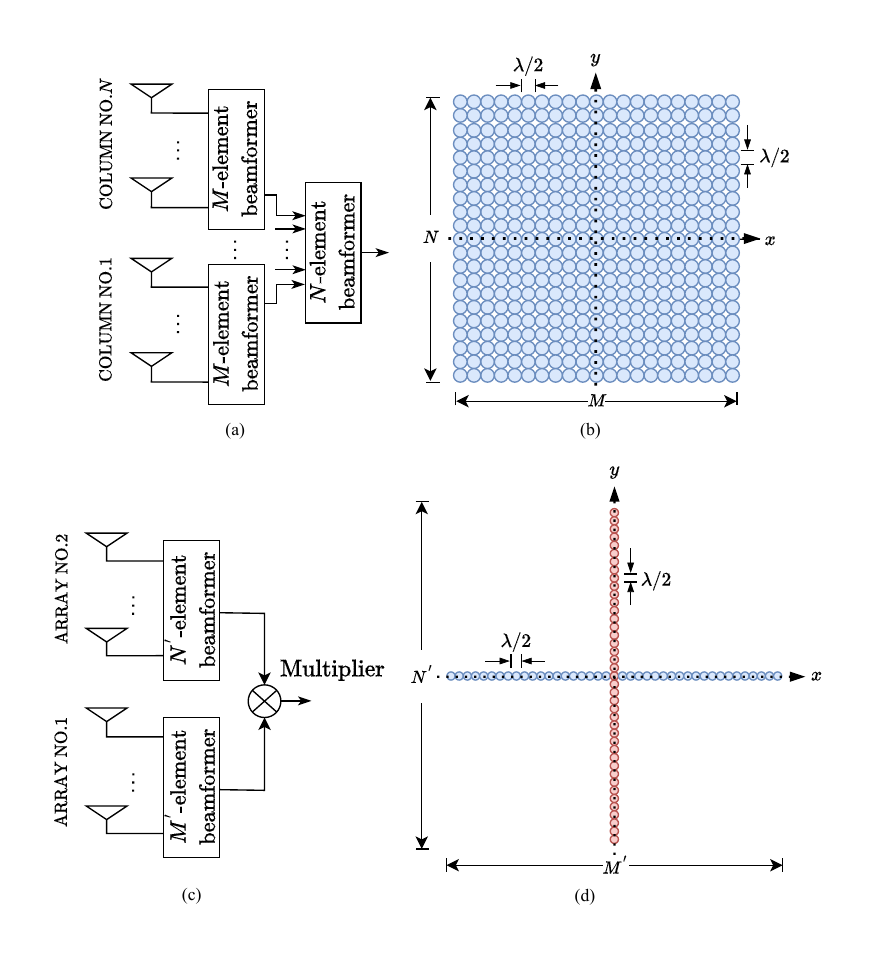}
			\caption{ (a) Architecture of the URA. (b) Distribution of array element in the URA. (c) Architecture of the MA. (d) Distribution of array element in the MA.}
			\label{f1}
		\end{figure}

		\subsection{Multiplicative Array for Pattern Synthesis}
		Consider a URA consisting of $M \times N$ isotropic elements defined in Cartesian coordinates, as shown in Fig. \ref{f1} (b). We choose the center of array as the phase reference point and each antenna element in the URA is denoted by $(m,n)$, where $m \in [-(M-1)/2,(M-1)/2] \ \text{and} \ n \in [-(N-1)/2,(N-1)/2 ]$ \footnote{Note that we assume the values of $M$ and $N$ are both odd numbers.}. The array factor $\mathbf{D}_{\scriptscriptstyle{URA}}= \{  D_{\scriptscriptstyle{URA}}^{\theta,\phi}  \} \in \mathbb{C}^{S_{\theta} \times S_{\phi}}$ in the $x\text{-}y$ plane can be defined as 
		\begin{align}
			&D_{\scriptscriptstyle{URA}}^{\theta,\phi} = \notag \\ & \frac{1}{MN}  \sum_{m= -\frac{M-1}{2}}^{\frac{M-1}{2}} \sum_{n= -\frac{N-1}{2}}^{\frac{N-1}{2}} {\rm{e}}^{ {\rm{j}} \frac{2 \pi}{\lambda} (m d_x {\rm{sin}}\theta {\rm{cos}}\phi +n d_y {\rm{sin}}\theta {\rm{sin}}\phi ) }  \cdot A_{\scriptscriptstyle{URA}}^{m,n}
			\label{AF_URA}
		\end{align}
		where $\theta$ and $\phi$ are the elevation angle and azimuth angle, respectively. $S_{\theta}$ and $S_{\phi}$ are the corresponding sampling number on elevation angle and azimuth angle. $\mathbf{A}_{\scriptscriptstyle{URA}}= \{  A_{\scriptscriptstyle{URA}}^{m,n}  \} \in \mathbb{C}^{M \times N}$ is the excitation matrix. $\lambda$ is the wavelength.  $d_x$ and $d_y$ are the element spacing between antenna elements in $x\text{-}$ and $y\text{-}$axes, which should not exceed a half-wavelength to prevent the occurrence of undesired grating lobes.

		We can map the given $x\text{-}y$ plane to $(u,v)$ space, which is a spherical coordinate system used to simplify and normalize the spatial direction. The mapping relationship with $x\text{-}y$ plane can be given \cite{abbasi2018compressive} by 
		\begin{align}
			u & = {\rm{sin}}\theta {\rm{cos}}\phi, \notag \\
			v & = {\rm{sin}}\theta {\rm{sin}}\phi.
			\label{eq10}
		\end{align}
		
		To facilitate subsequent analysis, we redefine the array factor of the URA in $(u,v)$ space as $\mathbf{D}_{\scriptscriptstyle{URA}}= \{  D_{\scriptscriptstyle{URA}}^{u,v}  \} \in \mathbb{C}^{U \times V}$, where $U$ and $V$ are the number of horizontal and vertical coordinates in the$(u,v)$ space both ranged in [-1,1], corresponding to the range of $\theta \in [0^{\circ}, 90^{\circ}]$ and $\phi \in [0^{\circ}, 360^{\circ}]$ in the $x\text{-}y$ plane. The previous definition of the URA will no longer be used. We have \cite{brandwood2012fourier}
		\begin{equation}
			\mathbf{D}_{\scriptscriptstyle{URA}} = \mathcal{F}_{u}^{-1}\mathcal{F}_{v}^{-1} (\mathbf{A}_{\scriptscriptstyle{URA}}) ,
			\label{eq1}
		\end{equation}
		where $\mathcal{F}_{u}^{-1}\mathcal{F}_{v}^{-1}$ denotes the two dimensional inverse Fourier transform (IFT) on $u$ and $v$ dimensions. We can rewrite the Eq. (\ref{AF_URA}) as
		\begin{equation}
			D_{\scriptscriptstyle{URA}}^{u,v} =\frac{1}{MN} \sum_{m= -\frac{M-1}{2}}^{\frac{M-1}{2}} \sum_{n= -\frac{N-1}{2}}^{\frac{N-1}{2}} {\rm e}^{{\rm j} \frac{2 \pi}{\lambda} (md_x u +  nd_y v)} \cdot A_{\scriptscriptstyle{URA}}^{m,n}.
			\label{AF_URA_UV}
		\end{equation}
		
		In fact, the synthesis technique described in following content requires $\mathbf{A}_{\scriptscriptstyle{URA}}$ to be separable, that is
		\begin{equation}
			\mathbf{A}_{\scriptscriptstyle{URA}}= \mathbf{w}_{\scriptscriptstyle{URA}}^{x} \cdot (\mathbf{w}_{\scriptscriptstyle{URA}}^{y})^{T},
			\label{A}
		\end{equation}
		where $\mathbf{w}_{\scriptscriptstyle{URA}}^x= \{  w_{\scriptscriptstyle{URA}}^{x,m}  \} \in \mathbb{C}^{M \times 1}, \mathbf{w}_{\scriptscriptstyle{URA}}^y= \{  w_{\scriptscriptstyle{URA}}^{y,n}  \} \in \mathbb{C}^{N \times 1}$ denotes the excitation vectors of the URA along $x\text{-}$ and $y\text{-}$ axes, respectively.  $()^{T}$ denotes a transpose operation. Therefore, by applying the Fourier transform separability property \cite{macphie2007mills}, the array factor can be rewritten as
		\begin{equation}
			\mathbf{D}_{\scriptscriptstyle{URA}} = \mathcal{F}_{u}^{-1}\mathcal{F}_{v}^{-1} (\mathbf{A}_{\scriptscriptstyle{URA}}) = \mathcal{F}_{u}^{-1}(\mathbf{w}_{\scriptscriptstyle{URA}}^{x}) \cdot \mathcal{F}_{v}^{-1}[(\mathbf{w}_{\scriptscriptstyle{URA}}^{y})^{T}].
			\label{AF_URA_separ}
		\end{equation}
		
		The power pattern $\mathbf{P}_{\scriptscriptstyle{URA}} \in \mathbb{C}^{M \times N}$ of the URA can be given by 
		\begin{equation}
			\mathbf{P}_{\scriptscriptstyle{URA}}  = \mathbf{D}_{\scriptscriptstyle{URA}} \times \mathbf{D}_{\scriptscriptstyle{URA}}^{*},
			\label{PP_1}
		\end{equation}
		where $(\times)$ denote the element wise (or Hadamard multiplication \cite{roberts1987digital}), and  $(\cdot)^*$ denotes the complex conjugate. Take Eq. (\ref{AF_URA_separ}) into Eq. (\ref{PP_1}), we have
		\begin{align}
			\mathbf{P}_{\scriptscriptstyle{URA}} = \{ \mathcal{F}_{u}^{-1}&(\mathbf{w}^x_{\scriptscriptstyle{URA}})  \cdot  \mathcal{F}_{v}^{-1}(\mathbf{w}^y_{\scriptscriptstyle{URA}})^{T} \}  \notag \\ & \times \{ \mathcal{F}_{u}^{-1}(\mathbf{w}^x_{\scriptscriptstyle{URA}}) \cdot  \mathcal{F}_{v}^{-1}(\mathbf{w}^y_{\scriptscriptstyle{URA}})^{T} \}^{*},
			\label{PP_2}
		\end{align}
		
	    By applying the distributive property of a Hadamard product, it can be rewritten as
		\begin{align}
			\mathbf{P}_{\scriptscriptstyle{URA}} =  \{ \mathcal{F}_{u}^{-1}&(\mathbf{w}^x_{\scriptscriptstyle{URA}}) \times [ \mathcal{F}_{u}^{-1}(\mathbf{w}^x_{\scriptscriptstyle{URA}})]^{*} \} \notag \\  &\cdot \{ \mathcal{F}_{v}^{-1}(\mathbf{w}^y_{\scriptscriptstyle{URA}})^{T} \times [ \mathcal{F}_{v}^{-1}(\mathbf{w}^y_{\scriptscriptstyle{URA}})^{T}]^{*}\}.
			\label{PP_3}
		\end{align}	
	
		According to Eq. (\ref{PP_3}), the power pattern of URA with separable excitation can be expressed as two multiplied antenna patterns by substituting the multiplication with auto-convolution of complex excitation vectors.
		
		\begin{figure}[!t]
			\centering
			\includegraphics[width=0.5\textwidth]{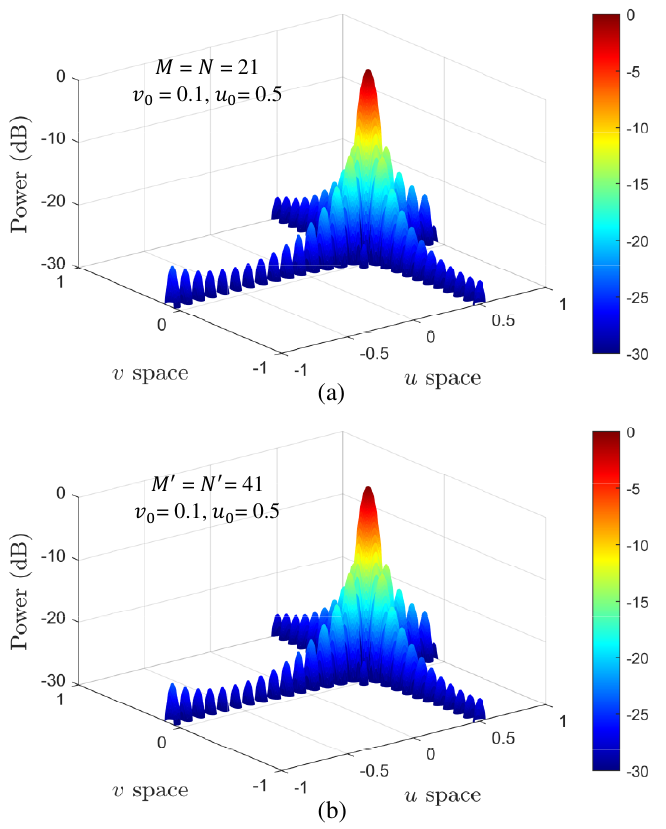}
			\caption{ (a) The power pattern of the URA. (b) The power pattern of the MA.}
			\label{Powerpattern}
		\end{figure}

	    Consider a MA configuration wherein each of the two orthogonal sub-arrays consists of $M'$ and $N'$ elements, as shown Fig. \ref{f1} (c) and (d). The antenna element in each sub-array of MA are denoted by $(m^{'},0)$ and $(0,n^{'})$ separately, with $m^{'} \in [-(M^{'}-1)/2,(M^{'}-1)/2] \ \text{and} \ n^{'} \in [-(N^{'}-1)/2,(N^{'}-1)/2 ]$. The power pattern $\mathbf{P}_{\scriptscriptstyle{MA}} \in \mathbb{C}^{U \times V}$ of MA is defined \cite{aumann2010pattern} by 
	 	\begin{equation}
			\mathbf{P}_{\scriptscriptstyle{MA}} = \mathbf{D}_{\scriptscriptstyle{MA}}^x  \cdot (\mathbf{D}_{\scriptscriptstyle{MA}}^{y})^{T},	
			\label{eq5}
		\end{equation}
		where $\mathbf{D}_{\scriptscriptstyle{MA}}^x$ and $\mathbf{D}_{\scriptscriptstyle{MA}}^y$ are the array pattern of the corresponding sub-arrays of the MA along $x\text{-}$ and $y\text{-}$ axes, with excitation vectors $\mathbf{w}^x_{\scriptscriptstyle{MA}} = \{ w^{x,m'}_{\scriptscriptstyle{MA}} \} \in \mathbb{C}^{M' \times 1}$ and $\mathbf{w}^y_{\scriptscriptstyle{MA}} = \{ w^{y,n'}_{\scriptscriptstyle{MA}} \} \in \mathbb{C}^{M' \times 1}$, which are designed to be auto-convolution of $\mathbf{w}^x_{\scriptscriptstyle{URA}}$ and $\mathbf{w}^x_{\scriptscriptstyle{URA}}$.
		\begin{align}
			\mathbf{w}^x_{\scriptscriptstyle{MA}}& = \mathbf{w}^x_{\scriptscriptstyle{URA}} \otimes \mathbf{w}^x_{\scriptscriptstyle{URA}},  \notag \\
			\mathbf{w}^y_{\scriptscriptstyle{MA}}& = \mathbf{w}^y_{\scriptscriptstyle{URA}} \otimes \mathbf{w}^y_{\scriptscriptstyle{URA}},  
			\label{auto-conv}
		\end{align}
		where $M^{'} = 2M-1, \ N^{'}=2N-1$ and $\otimes$ denotes the convolution operation. Substitute Eq. (\ref{auto-conv}) to Eq. (\ref{eq5}), we have
		\begin{align}
			\mathbf{P}_{\scriptscriptstyle{MA}} =  \mathcal{F}_{u}^{-1}(\mathbf{w}^x_{\scriptscriptstyle{URA}} \otimes \mathbf{w}^x_{\scriptscriptstyle{URA}}  ) \cdot  \mathcal{F}_{v}^{-1}[(\mathbf{w}^y_{\scriptscriptstyle{URA}})^{T} \otimes (\mathbf{w}^y_{\scriptscriptstyle{URA}})^{T}  ].
			\label{eq7}
		\end{align}
		
		By applying the Fourier convolution theorem \cite{macphie2007mills}, Eq. (\ref{eq7}) can be written as 
		\begin{align}
			\mathbf{P}_{\scriptscriptstyle{MA}}=  \{ \mathcal{F}_{u}^{-1}&(\mathbf{w}^x_{\scriptscriptstyle{URA}}) \times  \mathcal{F}_{u}^{-1}(\mathbf{w}^x_{\scriptscriptstyle{URA}}) \}  \notag \\ &\cdot \{ \mathcal{F}_{v}^{-1}(\mathbf{w}^y_{\scriptscriptstyle{URA}})^{T} \times  \mathcal{F}_{v}^{-1}(\mathbf{w}^y_{\scriptscriptstyle{URA}})^{T}\}.
			\label{PP_MA}
		\end{align}
		
		While excitation functions should satisfy the conjugate symmetry condition \cite{gloub1996matrix}, Eq. (\ref{PP_3}) and Eq. (\ref{PP_MA}) will be the same.
		\begin{equation}
			(\mathbf{w}^{x}_{\scriptscriptstyle{URA}})^{*} = \mathbf{w}^{x}_{\scriptscriptstyle{URA}} \cdot \mathbf{J} \ \text{and} \ (\mathbf{w}^{y}_{\scriptscriptstyle{URA}})^{*} = \mathbf{w}^{y}_{\scriptscriptstyle{URA}} \cdot \mathbf{J},
			\label{eq8}
		\end{equation}
		where $\mathbf{J}$ is a per-symmetric identity matrix. Note that all conventional array tapering and beam steering operations can be used for the MA and the URA the same way. 
		
		An example involving a URA of $M=N=21$ and a MA of $M^{'}=N^{'}=41$ isotropic elements with $\lambda /2$ element spacing is shown in Fig. \ref{f1}, respectively. The beam is steered to $(v_0=0.1,u_0=0.5)$ (i.e. approximately 11.3$^{\circ}$ in the azimuth and 30.8$^{\circ}$ in the elevation). The corresponding power pattern of the URA and the MA are shown in Fig. \ref{Powerpattern}. We can observe that URA have 441 elements and the corresponding MA requires only 82 elements to reach the same power pattern.

		This approach can be utilized to reduce the number of elements in a 2-D array from $M \times N$ to two single separate orthogonal arrays comprising $(2M-1) + (2N-1) $ elements, as suggested by the 1-D convolution. However, it is essential to note that the above proof is based on the analysis of narrowband assumption for antenna pattern synthesis applications. The MA concept has demonstrated its potential to significantly reduce the number of antenna elements in the pattern synthesis. Inspired by this and the strong need in channel sounding community to reduce the number of required antennas. In the following, we will expand the application of the MA into channel sounding and extend our discussion to include the real-world multi-path scenario.

		\subsection{Multiplicative Array for Channel Sounding} %
		To date, application of the MA for channel sounding has not been reported in the literature, to our best knowledge. Previous work has already demonstrated the effectiveness of MA's performance on array thinning and pattern synthesis, and its advantages will become even more significant as the array size grows. Hence, in this work we introduce the MA into channel sounding, with an effort to replace the URA which requires a demanding number of antennas for channel sounding. 

		Consider the URA and the MA in multi-path scenario, with the same array configuration as previously. The direction of $k \text{-th}$ incident path is determined by the elevation and azimuth angles $\theta_k$ and $\phi_k$ respectively with $k \in [1,K]$. The elements in both arrays are placed in the same $x\text{-}y$ plane with the identical element spacing of $d$ in both directions of $x$ and $y$. 
		
		The channel frequency response (CFR) at the element denoted by $(m,n)$ in URA can be given \cite{balanis2016antenna} by
		\begin{align}
			H_{\scriptscriptstyle{URA}}^{m,n}(f) = \sum_{k=1}^{K} \alpha_k \cdot {\rm e}^{-{\rm{j}}2 \pi f \tau_{k}} \cdot {\rm e}^{{\rm j} \frac{2 \pi}{\lambda} (md_x u_k +  nd_y v_k)}, 
			\label{CFR_URA}
		\end{align}
		where $\alpha_k$ and $\tau_k$ denote the complex amplitude and delay of the $k \text{-th}$ path, respectively. 
		
		Furthermore, according to the special architecture of the MA in Fig. \ref{f1} (c), the CFR of elements denoted by $(m',0)$ and $(0,n')$ in each sub-array of the MA along $x\text{-}$ and $y\text{-}$axes can be given by
		\begin{align}
			H^{{x,m'}}_{\scriptscriptstyle{MA}}(f) &= \sum_{k=1}^{K} \alpha_k \cdot {\rm{e}}^{-{\rm{j}}2 \pi f \tau_k} \cdot {\rm{e}}^{{\rm{j}} \frac{2 \pi}{\lambda} (m' d u_k) } \notag \\
			H^{{y,n'}}_{\scriptscriptstyle{MA}}(f) &= \sum_{k=1}^{K} \alpha_k \cdot  {\rm{e}}^{-{\rm{j}}2 \pi f \tau_k} \cdot {\rm{e}}^{{\rm{j}} \frac{2 \pi}{\lambda} (n^{'} d v_k) }.
			\label{CFR_MA}
		\end{align}
		
		We aim to estimate the path parameters $\{\alpha_k, \theta_{k}, \phi_k, \tau_k\}$ by applying the MA in channel sounding under multi-path scenario. Note that the performance of the MA has been proven to be the same as the URA, considering only one-path scenario (i.e. the pattern synthesis). Hence, note that we employ the classical beamforming (CBF) in both URA and MA to discover the characteristic of MA.
		
		Using the CBF under plane-wave assumption, the beam pattern of the URA can be derived by coherently summing the responses of individual elements as follows:
		\begin{align}
			&B_{\scriptscriptstyle{URA}}^{u,v}(f,\theta,\phi)   \notag \\ &= \frac{1}{MN} \sum_{m=- \frac{M-1}{2}}^{\frac{M-1}{2}} \sum_{n=- \frac{N-1}{2}}^{\frac{N-1}{2}} A_{\scriptscriptstyle{URA}}^{m,n}(f, \theta, \phi) \cdot  H_{\scriptscriptstyle{URA}}^{m,n}(f) \notag \\
			&=   \frac{1}{MN} \sum_{k=1}^{K} \sum_{m=- \frac{M-1}{2}}^{\frac{M-1}{2}} \sum_{n=- \frac{N-1}{2}}^{\frac{N-1}{2}} \alpha_k \cdot {\rm{e}}^{{\rm{j}} \frac{2 \pi }{\lambda} \big[  md(u- u_k)} \notag \\ & \ \ \ \ ^{+ nd(v-v_k) \big]} \cdot {\rm{e}}^{-{\rm{j}} 2 \pi f \tau_k} \notag \\ 
			& =  \sum_{k=1}^{K} \alpha_k \cdot {\rm{e}}^{-{\rm{j}} 2 \pi f \tau_k} \cdot g_{\scriptscriptstyle{URA}}^k(\theta,\phi) 
			\label{beam_URA}
		\end{align}
	
		where $ g_{\scriptscriptstyle{URA}}^k(\theta,\phi) $ denotes the unit URA beam pattern term of the $k\text{-th}$ path with CBF. The peak location of $|g_{\scriptscriptstyle{URA}}^k(\theta,\phi)|$ gives the initial estimation of angular parameters $\{ \theta_k , \phi_k \}$. Then, the parameters $\{ \alpha_k , \tau_k \}$ can be obtained by performing IFT of $B_{\scriptscriptstyle{URA}}^{u,v}(f,\theta,\phi) $ at $\theta = \theta_k$ which corresponds to the power angular delay spectrum (PADP) $b_{\scriptscriptstyle{URA}}(\tau,\theta,\phi)$ of URA for $k \text{-th}$ path. According to Eq. (\ref{beam_URA}), the PADP of URA can be given by
		\begin{align}
			&b_{\scriptscriptstyle{URA}}(\tau,\theta,\phi) \notag \\ &=  \sum_{f=f_1}^{f_L} B_{\scriptscriptstyle{URA}}(f,\theta,\phi) \cdot {\rm{e}}^{{\rm{j}} 2 \pi f \tau} \notag \\ &= \sum_{k=1}^{K} \alpha_k \cdot \delta(\tau-\tau_k)  \cdot g_{\scriptscriptstyle{URA}}^k(\theta,\phi) 
			\label{PADP_URA}
		\end{align}
		where $L$ denotes the number of frequency points, with frequency ranges from frequency points from  $f_1$ to $f_L$. 
		
		Subsequently, on the basis of Eq. (\ref{CFR_MA}) and (\ref{beam_URA}), the beam pattern $B_{\scriptscriptstyle{MA}}(f,\theta,\phi)$ of the MA can be given by

		\begin{align}
			&B_{\scriptscriptstyle{MA}}(f,\theta,\phi) \notag \\  =& \bigg[ \frac{1}{M} \sum_{m^{'}=- \frac{M^{'}-1}{2}}^{\frac{M^{'}-1}{2}} w^{x,m'}_{\scriptscriptstyle{MA}}(f, \theta, \phi) \cdot H^{{x,m'}}_{\scriptscriptstyle{MA}}(f) \bigg] \notag \\ & \cdot \bigg[ \frac{1}{N} \sum_{n^{'}=- \frac{N^{'}-1}{2}}^{\frac{N^{'}-1}{2}} w^{y,n'}_{\scriptscriptstyle{MA}} (f, \theta, \phi) \cdot H^{{y,n'}}_{\scriptscriptstyle{MA}}(f)\bigg] \notag \\ =&  \bigg[  \sum_{k=1}^{K}  \alpha_k\cdot {\rm{e}}^{-{\rm{j}} 2 \pi f \tau_k} \cdot g_{\scriptscriptstyle{MA}}^{x,k}(\theta,\phi)  \bigg] \notag \\ & \cdot  \bigg[  \sum_{k=1}^{K}   \alpha_k \cdot{\rm{e}}^{-{\rm{j}} 2 \pi f \tau_k} \cdot g_{\scriptscriptstyle{MA}}^{y,k}(\theta,\phi) \bigg]  
			\label{beam_MA}
		\end{align}
		where $g_{\scriptscriptstyle{MA}}^{x,k}(\theta,\phi)$ and $g_{\scriptscriptstyle{MA}}^{y,k}(\theta,\phi)$ are the unit beam pattern of each sub-array of the MA along $x\text{-}$ and $y\text{-}$ axes, respectively. By performing IFT on Eq. (\ref{beam_MA}), we can obtain the PADP of the MA which can be given by
		\begin{align}
			&b_{\scriptscriptstyle{MA}}(\tau,\theta,\phi) \notag \\&=  \sum_{f=f_1}^{f_L} B_{\scriptscriptstyle{MA}}(f, \theta, \phi) \cdot {\rm{e}}^{{\rm{j}} 2 \pi f \tau} \notag \\ &= \bigg[ \sum_{k=1}^{K}  \alpha_k \cdot \delta(\tau-\tau_k) \cdot g_{\scriptscriptstyle{MA}}^{x,k}(\theta,\phi)  \bigg] \notag \\ &\cdot \bigg[  \sum_{k=1}^{K}  \alpha_k \cdot \delta(\tau-\tau_k) \cdot g_{\scriptscriptstyle{MA}}^{y,k}(\theta,\phi)  \bigg]. 
			\label{PADP_MA}
		\end{align}
		
		\begin{figure}[!t]
			\centering
			\includegraphics[width=0.52\textwidth]{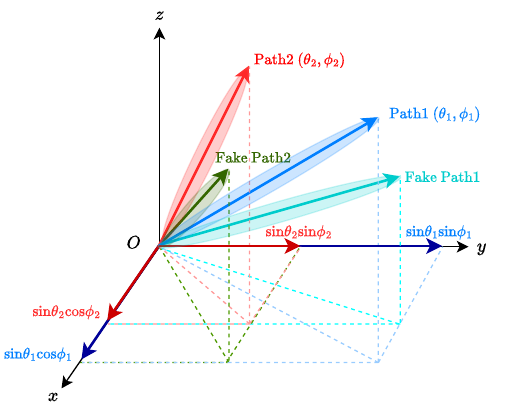}
			\caption{ An illustration of the occurrence of fake paths with two incident paths.}
			\label{num_fake}
		\end{figure}

		The multiplication of unit beam pattern $g_{\scriptscriptstyle{MA}}^{x,k}(\theta,\phi)$ and $g_{\scriptscriptstyle{MA}}^{y,k}(\theta,\phi)$ can be used to estimate the angular parameters. However, above estimations can only be realized under one-path scenario, the angular estimation results of Eq. (\ref{beam_URA}) and (\ref{beam_MA}) will be the same for $K=1$, which will not introduce extra phasor terms after the multiplication operation. Consider the multi-path scenario with $K > 1$, the $k\text{-th}$ path with incident elevation angle $\theta_k$ and azimuth angle $\phi_k$ will introduce a phasor term on two sub-array patterns, respectively. As a result, $K$ incident paths will introduce $K$ phasor terms on each sub-array pattern and the multiplication of these two sub-array pattern will result in $K^{2}$ combinations of phasor terms. The extra $(K^2 - K)$ items are all the fake paths caused by cross multiplication of each two true paths' phasor terms among all the incident paths, and their angular positions and delays are all disordered. 	
		
		Below, we will demotrate that, the angular and delay profiles, though distorted for the MA, will present fixed patterns compared to the URA. This phenomenon will be essential for applying algorithms to retrieve the true angle and delay patterns, as shown later.

		\begin{figure}[!t]
			\centering
			\includegraphics[width=0.42\textwidth]{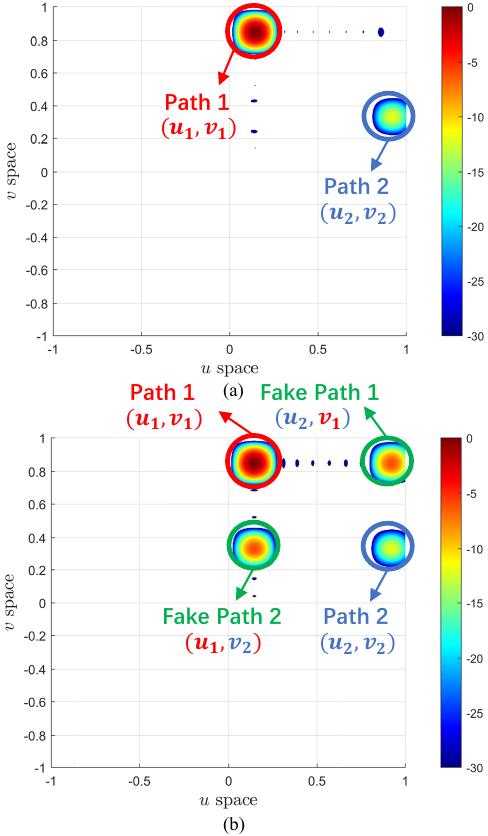}
			\caption{(a) The power pattern of the URA with two paths in $(u,v)$ space. (b) The power pattern of the MA with two paths in $(u,v)$ space.}
			\label{angualr_space}
		\end{figure}
		
		\subsection{Numerical simulations} %

		\begin{itemize}
		\item[1)] Power angle profile analysis for the MA
		\end{itemize}
		
		An intuitive illustration of the occurrence of fake paths with two incident paths is shown as Fig. \ref{num_fake}. We can observe that each of two true paths will introduce a pair of phasor terms $\{ {\rm{sin}}\theta {\rm{cos}}\phi, {\rm{sin}}\theta {\rm{sin}}\phi \}$ on the $x\text{-}$ and the $y\text{-}$ axis, respectively. The multiplication operation will mix these two pairs of phasor terms, resulting in four pairs of phasor terms which are $\{ {\rm{sin}}\theta_1 {\rm{cos}}\phi_1, {\rm{sin}}\theta_1 {\rm{sin}}\phi_1 \}$, $\{ {\rm{sin}}\theta_2 {\rm{cos}}\phi_2, {\rm{sin}}\theta_2 {\rm{sin}}\phi_2 \}$, $\{ {\rm{sin}}\theta_1 {\rm{cos}}\phi_1, {\rm{sin}}\theta_2 {\rm{sin}}\phi_2 \}$, and $\{ {\rm{sin}}\theta_2 {\rm{cos}}\phi_2, {\rm{sin}}\theta_1 {\rm{sin}}\phi_1 \}$. The first two pairs are the correct phasor terms for two true incident paths, while the other two are fake.

		\begin{figure}[!t]
			\centering
			\includegraphics[width=0.43\textwidth]{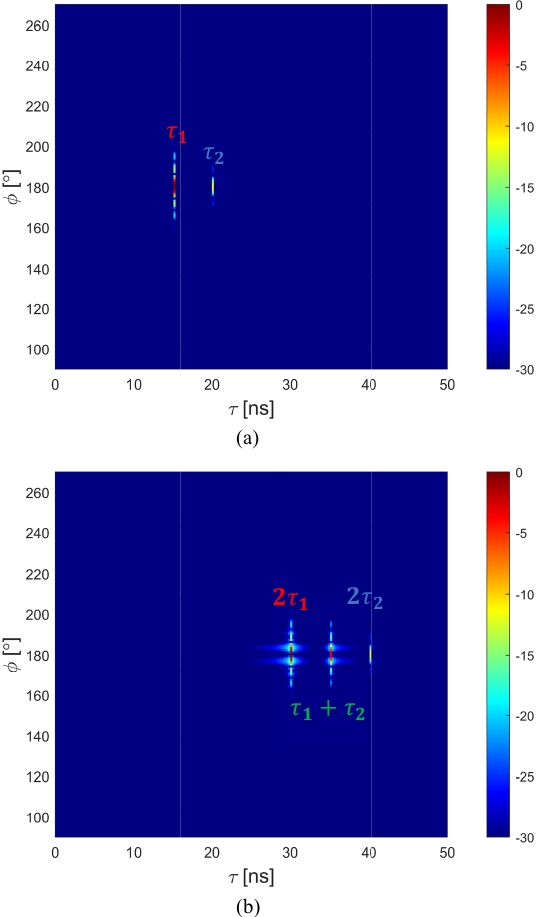}
			\caption{(a) The PADP of the URA with two paths. (b) The PADP of the MA with two paths.}
			\label{delay_space}
		\end{figure}
		
		To prove the correctness of above formulations, a simulation is conducted where the results are shown in Fig. \ref{angualr_space}. We define a URA consisting of $ 21 \times 21 $ isotropic elements with half-wavelength element spacing and a corresponding MA with $41$ elements on each sub-array. Besides, we apply a 30-dB Chebyshev taper in the $x\text{-}$direction and $y\text{-}$direction. Consider two incident paths with the parameters of $\{\alpha_1 = 0\ \text{dB}, \theta_1 = 60^{\circ}, \phi_1=80^{\circ} \}$ and $\{\alpha_2 = -12\ \text{dB}, \theta_2 = 80^{\circ}, \phi_2=20^{\circ} \}$. We evaluate the power pattern of both URA and MA according to Eq. (\ref{eq5}) and Eq. (\ref{eq7}) in $(u,v)$ space. Therefore, the angular parameters of two incident paths in $(u,v)$ space can be obtained by $(u_1={\rm{sin}}60^{ \circ } {\rm{cos}}80^{ \circ } \approx 0.15, v_1={\rm{sin}}60^{ \circ } {\rm{sin}}80^{ \circ } \approx 0.85 )$ and $(u_2={\rm{sin}}80^{ \circ } {\rm{cos}}20^{ \circ } \approx 0.92, v_2={\rm{sin}}80^{ \circ } {\rm{sin}}20^{ \circ } \approx 0.33 )$, and these two pairs of angular positions are corresponding to the results in Fig. \ref{angualr_space} (a) which only shows two peak locations with power values of approximate $0\  \text{dB}$ and $-12\ \text{dB}$, respectively. Furthermore, by comparing the Fig. \ref{angualr_space} (a) and (b), it can be noticed that the MA yields two extra fake paths with angular positions $(u_1,v_2)$ and $(u_2,v_1)$. The power of each fake path is around $-6 \ \text{dB}$, which equals to the value of $\sqrt{\alpha_1 \alpha_2}$. The other two paths remained the same angular positions and powers as URA's power pattern are the true paths. These simulation results validate Eq. (\ref{beam_URA}) and Eq. (\ref{beam_MA}) derived above.

		\begin{itemize}
			\item[2)] Power delay profile analysis for the MA
		\end{itemize}

		However, above formulations and simulations only demonstrate the performance of the MA in estimating angular parameters, considering the scenario of multi-path with single frequency point. The performance of the MA in delay domain
		will be discussed in detail in the following, considering the multi-path scenario in wideband from $f_1$ to $f_L$ with multi frequency points.
		
		Through analysis and comparison of the Eq. (\ref{PADP_URA}) and Eq. (\ref{PADP_MA}), the PADP of the MA also generates $K^{2}$ terms which has similar principle with Eq. (\ref{beam_MA}). The IFT operation compensates the phasor terms brought by delay of each path, either true or fake. The delay of true paths will be doubled, and the delay of each fake path is the sum of each two true paths' delay whose multiplication introduces this fake path.
		
		In order to better illustrate this phenomenon, we rewrite Eq. (\ref{PADP_MA}) based on an example, which involves a MA subjected to two different incident paths with parameter $(\alpha_1, \theta_1, \phi_1, \tau_1) \ \text{and} \ (\alpha_2, \theta_2, \phi_2, \tau_2) $. Hence, Eq. (\ref{PADP_MA}) can be rewritten as
	
		
		\begin{flalign}
			&b_{\scriptscriptstyle{MA}}(\tau,\theta,\phi) \notag \\ &= \bigg[  \alpha_1^2 \cdot \delta(\tau-2\tau_1) \cdot g_{\scriptscriptstyle{MA}}^{x,1}(\theta,\phi) \cdot g_{\scriptscriptstyle{MA}}^{y,1}(\theta,\phi)  \bigg] \notag \\ &+ \bigg[  \alpha_2^2 \cdot \delta(\tau-2\tau_2) \cdot g_{\scriptscriptstyle{MA}}^{x,2}(\theta,\phi) \cdot g_{\scriptscriptstyle{MA}}^{y,2}(\theta,\phi)  \bigg] \notag \\ &+ \bigg[  \alpha_1\alpha_2 \cdot \delta(\tau-(\tau_1+\tau_2)) \cdot g_{\scriptscriptstyle{MA}}^{x,1}(\theta,\phi) \cdot g_{\scriptscriptstyle{MA}}^{y,2}(\theta,\phi)  \bigg]  \notag \\ &+ \bigg[  \alpha_1\alpha_2 \cdot \delta(\tau-(\tau_1+\tau_2)) \cdot g_{\scriptscriptstyle{MA}}^{x,2}(\theta,\phi) \cdot g_{\scriptscriptstyle{MA}}^{y,1}(\theta,\phi)  \bigg],
			\label{PADP_MA_simu}
		\end{flalign}
		which reveals that the beam is steered towards four paths with the parameters $(\alpha_1, u_1,v_1,2 \tau_1)$, $(\alpha_2, u_2,v_2,2 \tau_2)$, $(\sqrt{\alpha_1 \alpha_2} ,u_1,v_2,(\tau_1+\tau_2))$ and $(\sqrt{\alpha_1 \alpha_2}, u_2,v_1,(\tau_1+\tau_2))$. Among these paths, two have the correct angular parameters but with double delay, and the others are additional fake paths introduced by multiplication between the beam pattern of sub-arrays.

		We conduct a simulation to support the phenomena demonstrated by the aforementioned equations, shown as Fig. \ref{delay_space}. Consider a wideband scenario with 4 GHz bandwidth from 26 GHz to 30 GHz to identify the multi-path components in delay domain. The array sizes of URA and MA keep the same as previous simulation configuration, and also subjected to two incident paths. We assume the elevation angles $\theta_{k} $ and azimuth angles $\phi_k$ for two incident paths are both $90^{\circ }$ and $180^{\circ }$, respectively. Two paths have the different delays of 15 ns and 20 ns with powers of $0\ \text{dB}$ and $-12 \ \text{dB}$, respectively. The scan azimuth angle for URA and MA is from 90$^{\circ}$ to 270$^{\circ}$.
		
		By observing Fig. \ref{delay_space} (a), the PADP of the URA shows two peak locations at $\phi=180^{\circ}$ with delay of 15ns and 20ns, corresponding to the given parameters. Compared with Fig. \ref{delay_space} (b), the PADP of the MA detects three peak locations with delay of 30 ns, 35 ns and 40 ns.	Through previous derived Eq. (\ref{PADP_MA}), we can determine that the true paths have been spread spectrum and delays are all doubled. The fake paths' delay are the sum of two true path's delays which leads to result of 35 ns. And these two fake paths overlap with each other, resulting one peak location is shown.
		
		As a summary, we introduce MA into channel sounding and evaluate its impact on angular and delay profiles considering multi-path scenarios. Notice that, it will yield extra fake paths and distortion for both angular and delay domain in wideband. As a result, the MA concept cannot be directly applied for channel sounding applications. However, as detailed in our work, the strucutre and pattern of the distorted delay and angle profiles of the MA can be well known. In this work, a novel algorithm is introduced for the channel sounding using the MA concept, as detailed below.

		\section{Proposed Solutions}
		In this section, a novel parameter estimation algorithm based on the SIC is introduced to address this problem. The core idea of this algorithm is to gradually remove the CFR of strongest path in each iteration and fake paths will vanish along the target path being removed.

		\subsection{Algorithm description}
		The array element response vector $\mathbf{H}^{x}_{\scriptscriptstyle{MA}} (f) \in \mathbb{C}^{M^{'} \times 1}$ and $\mathbf{H}^{y}_{\scriptscriptstyle{MA}} (f) \in \mathbb{C}^{N^{'} \times 1}$ of MA along $x\text{-}$ and $y\text{-}$ axes are defined as
		\begin{align}
			\mathbf{H}^{x}_{\scriptscriptstyle{MA}}(f) &= [H^{{x,1}}_{\scriptscriptstyle{MA}}(f),...,H^{{x,M'}}_{\scriptscriptstyle{MA}}(f) ]^{T}, \notag \\ \mathbf{H}^{y}_{\scriptscriptstyle{MA}}(f) &= [{H}^{{\rm{y}}}_{{\rm c},1}(f),...,{H}^{{\rm{y}}}_{{\rm c},N^{'}}(f) ]^{T},
			\label{HHH}
		\end{align}
		where $H^{{x,m'}}_{\scriptscriptstyle{MA}}(f)$ and $H^{{y,n'}}_{\scriptscriptstyle{MA}}(f)$ were defined is Eq. (\ref{CFR_MA}).
		
		For brief, we will substitute $\mathbf{H}_{\scriptscriptstyle{MA}}(f)$ for $\mathbf{H}^{x}_{\scriptscriptstyle{MA}}(f)$ and $\mathbf{H}^{y}_{\scriptscriptstyle{MA}}(f)$ as the combination of above two vector in the following discussion.
		\begin{align}
			\mathbf{H}_{\scriptscriptstyle{MA}}(f) = [\mathbf{H}^{x}_{\scriptscriptstyle{MA}}(f) ; \mathbf{H}^{y}_{\scriptscriptstyle{MA}}(f)] \in \mathbb{C}^{(M^{'}+N^{'}) \times 1}.
			\label{H}
		\end{align}

		\begin{algorithm}[t]
			\caption{Proposed algorithm}
			\SetKwInOut{Initialize}{Initialize}
			\SetKwInput{Output}{Output}
			\KwIn{$\mathbf{H}_{\scriptscriptstyle{MA}}(f)$, $f \in [f_1,f_L]$, $\epsilon $}
			\KwOut{{$(\hat{\alpha}_k, \hat{\tau}_k, \hat{\theta}_k, \hat{\phi}_k)$}, $k \in [1,K]$}
			\Initialize{ $\hat{{\alpha}}_{1} := 1$, $\hat{{\alpha}}_{max} := 1$,  $k:=1$, $q:=0$, maximum iteration number $K$ ;}
			
			$\mathbf{H}^{q}_{\scriptscriptstyle{MA}}(f) := \mathbf{H}_{\scriptscriptstyle{MA}}(f), \mathbf{h}^q_{\scriptscriptstyle{MA}}(\tau) := IFT(\mathbf{H}^{q}_{\scriptscriptstyle{MA}}(f)); $
			
			\While{$ \hat{\alpha}_{k} \textgreater \hat{\alpha}_{max} \cdot 10^{- \epsilon / 20 } $}
			{
				Apply Eq. (\ref{eq7}), (\ref{beam_MA}) on $\mathbf{H}^{q}_{\scriptscriptstyle{MA}}(f_{\rm center})$ to obtain the current beam pattern $B^q_{\scriptscriptstyle{MA}}(f_{\rm center},\theta,\phi)$, where $f_{\rm center}$ denotes the center frequency point. Then, detect the strongest path in beam pattern to capture the angular parameters for $k \text{-th}$ path $\{ \hat{\theta}_k, \hat{\phi}_k \} $;
				
				Apply Eq. (\ref{PADP_MA}) on $\mathbf{H}^{q}_{\scriptscriptstyle{MA}}(f)$ to obtain the PADP $b^{q}_{\scriptscriptstyle{MA}}(\tau, \phi) $ at $\theta = \hat{\theta}_k$. Detect the peak location in PADP to estimate the parameters $\{ \hat{\phi}_k, \hat{\tau}_k \}$, and update azimuth angle;
				
				Generate the label vector $s_k(\tau)$ to extract the single detected path's CIR $\mathbf {\hat{h}}_{{\scriptscriptstyle{MA}},k}(\tau)$ of $k \text{-th}$ path from $\mathbf{h}^{q}_{\scriptscriptstyle{MA}}(\tau)$;
				
				Reconstruct the CFR $ \mathbf {\hat{H}}_{{\scriptscriptstyle{MA}},k}(f) = FT(\mathbf {\hat{h}}_{{\scriptscriptstyle{MA}},k}(\tau))$, and perform Eq. (\ref{PADP_MA}) to estimate the whole parameters $(\hat{\alpha}_k, \hat{\tau}_k, \hat{\theta}_k, \hat{\phi}_k)$. Base on these parameters to reconstruct the CFR $\mathbf {\hat{H}^{'}}_{{\scriptscriptstyle{MA}},k}(f)$ for $k \text{-th}$ path.
				
				\If{k:=1} 
				{
					$\hat{{\alpha}}_{max}:=\hat{{\alpha}}_{k}$;
				}
				
				Remove the path's reconstructed CFR  $\mathbf {\hat{H}^{'}}_{{\scriptscriptstyle{MA}},k}(f)$ to update the CFR 
				$\mathbf{H}^{q+1}_{\scriptscriptstyle{MA}}(f)$ for next iteration, $\mathbf{H}^{q+1}_{\scriptscriptstyle{MA}}(f) = \mathbf{H}^{q}_{\scriptscriptstyle{MA}}(f) - \mathbf {\hat{H}^{'}}_{{\scriptscriptstyle{MA}},k}(f)$;
				
				$ k:=k+1, q:=q+1$;
				
			}
			
			\label{ag}
			
		\end{algorithm}

		In the following algorithm description, superscript and subscript numerals are appended to the CFR $\mathbf{H}_{\scriptscriptstyle{MA}}(f)$, beam pattern $B_{\scriptscriptstyle{MA}} (f,\theta,\phi)$ and PADP $b_{\scriptscriptstyle{MA}} (\tau, \phi) $ to denote that these terms need to be updated in each iteration. The superscript number $q$ denotes the $q$ path(s) removed. For example, $\mathbf{H}^{1}_{\scriptscriptstyle{MA}} (f)$ denotes the CFR with 1 path removed, and $b^2_{\scriptscriptstyle{MA}} (\tau, \phi) $ denotes the PADP with 2 paths removed. Initially, the CFR and PADP begin with the superscript number $q=0$. The subscript number $k$ denotes that there only contain $k\text{-th}$ path, such as $\mathbf {H}_{{\scriptscriptstyle{MA}},1}(f)$ represents the first detected path's CFR, $\mathbf {h}_{{\scriptscriptstyle{MA}},2}(\tau)$ represents the second detected path's CIR.

		\begin{figure}[!t]
			\centering
			\includegraphics[width=0.42\textwidth]{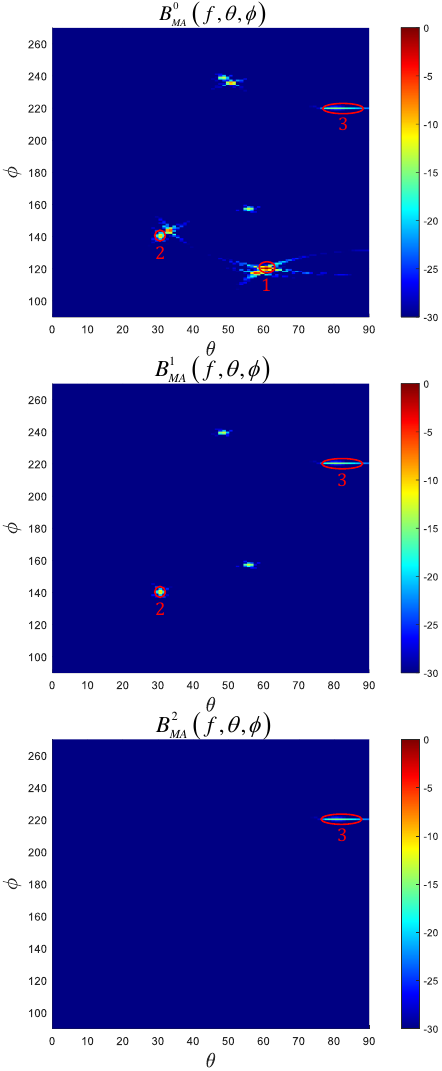}
			\caption{The beam pattern of $B^q_{\scriptscriptstyle{MA}}(f,\theta,\phi)$, $q \in [0,2]$, where the superscript denotes the $q$ path(s) are removed.}
			\label{BP}
		\end{figure}

		The procedure is detailed as below:
		\begin{itemize}
			\item [1.] 
			According to the initial CFR $\mathbf{H}^{0}_{\scriptscriptstyle{MA}}(f)$ for MA, we firstly apply IFT to obtain the CIR $\mathbf{h}^{0}_{\scriptscriptstyle{MA}}(\tau)$. Then, select the CFR based on center frequency point $\mathbf{H}_{\scriptscriptstyle{MA}}^{0}(f_{\rm center})$, and apply Eq. (\ref{eq7}) and Eq. (\ref{beam_MA}) to obtain the initial beam pattern $B^{0}_{\scriptscriptstyle{MA}}(f,\theta,\phi)$. Detect the strongest path in beam pattern to capture the angular position $\{ \hat{\theta}_1, \hat{\phi}_1 \} $. 
			
			\item [2.] 
			Apply Eq. (\ref{PADP_MA}) to obtain the current PADP $b^{0}_{\scriptscriptstyle{MA}} (\tau, \phi) $ at $\theta = \hat{\theta}_1$, and detect the peak location to estimate the azimuth angle and delay $\{ \hat{\phi}_1, \hat{\tau}_1 \}$ for path 1, where $\hat{\tau}_1$ takes half of estimated value. The azimuth angle will be updated in this iteration.
			\item [3.]
			Remove the CFR of path 1 from $\mathbf{H}^{0}_{\scriptscriptstyle{MA}}(f)$ to update the current CFR, obtaining $\mathbf{H}^{1}_{\scriptscriptstyle{MA}}(f)$. The detailed steps are shown as following.
           \begin{enumerate} 
			\item [3.1.]
			Base on above estimation results of elevation angle, azimuth angle and delay $( \hat{\theta_1}, \hat{\phi_1}, \hat{\tau}_1 $), we assume the amplitude $\hat{\alpha}_1 = 1$ for path 1, and reconstruct the CFR $\mathbf {H}_{{\scriptscriptstyle{MA}},1}(f)$. Subsequently, synthesis the CIR $\mathbf {h}_{{\scriptscriptstyle{MA}},1}(\tau)$ by performing IFT on $\mathbf {H}_{{\scriptscriptstyle{MA}},1}(f)$.  
			
	       \end{enumerate}
			
	       \begin{enumerate}	
			\item [3.2.]
			Generate one label vector $\mathbf{s}(\tau)$ with the same size as $\mathbf {h}_{{\scriptscriptstyle{MA}},1}(\tau)$, where for the $k \text{-th}$ iteration it can be given by
			\begin{equation}
				s_k(\tau)=\left\{
				\begin{aligned}
					1, &  \left| {h}_{{\scriptscriptstyle{MA}}, k} (\tau) \right| > \alpha_{k} \cdot 10 ^{\frac{-\epsilon}{20} }  \\ 0, & \ otherwise
				\end{aligned}
				\right.
				\label{eq18}
			\end{equation}
			
			where $\epsilon$ denotes the threshold value in decibel \cite{zhang2019near}. The label vector $\mathbf{s}(\tau)$ is aim to extract the single detected path's CIR used for later parameter estimation. Because we cannot directly estimate each path's power from PADP $b^{0}_{\scriptscriptstyle{MA}}(\tau, \phi) $, the existence of fake paths may cause aliasing and lead to error.
	        \end{enumerate}
			
		    \begin{enumerate}
			\item [3.3.]
			Remove the CIR of path 1 from the current CIR $\mathbf{h}^{0}_{\scriptscriptstyle{MA}} (\tau)$ and obtain the updated CIR $\mathbf {\hat{h}}_{{\scriptscriptstyle{MA}},1}(\tau)$ by
			\begin{align}
				\mathbf {\hat{h}}_{{\scriptscriptstyle{MA}},1}(\tau)  &= \mathbf{h}^{0}_{\scriptscriptstyle{MA}}(\tau) \times s_1(\tau).
				\label{eq19}
			\end{align}
			
  		    \end{enumerate}
		    
		    \begin{enumerate}
			\item [3.4.]
			Apply Fourier transform on $\mathbf {\hat{h}}_{{\scriptscriptstyle{MA}},1}(\tau) $ to obtain the CFR $\mathbf {\hat{H}}_{{\scriptscriptstyle{MA}},1}(f)$ for first path, and then perform the Eq. (\ref{PADP_MA}) on it to estimate and update the power $\hat{\alpha}_1$ by detecting the peak location. By now, all the parameters $(\hat{\alpha}_1, \hat{\tau}_1, \hat{\theta}_1, \hat{\phi}_1)$ for the first path have been estimated successfully. According to these parameters, we can reconstruct the CFR $\mathbf {\hat{H}^{'}}_{{\scriptscriptstyle{MA}},1}(f)$ for the first path. Here, we take the first path's power as the $\hat{{\alpha}}_{max}$ used for the judgment of later iteration. 
			\end{enumerate}
			
			\begin{enumerate}
			\item [3.5.]
			Remove the reconstructed CFR $\mathbf {\hat{H}^{'}}_{{\scriptscriptstyle{MA}},1}(f)$ of detected path from the initial CFR  $\mathbf{H}^{0}_{\scriptscriptstyle{MA}}(f)$ to update the current CFR $\mathbf{H}^{1}_{\scriptscriptstyle{MA}}(f)$ for next iteration.
			\begin{align}
				\mathbf{H}^{1}_{\scriptscriptstyle{MA}}(f) = \mathbf{H}^{0}_{\scriptscriptstyle{MA}}(f) - \mathbf {\hat{H}^{'}}_{{\scriptscriptstyle{MA}},1}(f).
				\label{eq20}
			\end{align}

			\end{enumerate}

			\item [4.]
			Repeat the above steps until the path's estimated power falls outside the predetermined dynamic range.

 		\end{itemize}

 		The whole procedure of the proposed algorithm is presented in Algorithm 1. According to Eq. (\ref{beam_MA}) and (\ref{PADP_MA}), the power of each extra fake paths is the multiplication of two true paths' power. It indicates that the fake path's power must be lower than one of the two true paths whose multiplication yields this fake path. Therefore, the path with strongest power must be the true path, and fake paths will vanish with the detected path removed. In the proposed algorithm, one strongest path's CFR will be removed for each iteration and the fake paths introduced by removed path will disappear, which will not interfere the detection of next strongest path. Therefore, all the target paths can be gradually detected with the power values ranked in descending order.

 		\subsection{Simulation Results}

 		In the simulation, we consider a MA consisting of $M^{'}=199$ and $N^{'}=199$ isotropic antennas with half-wavelength element spacing (which can mimic a URA of size 100$\times$100), distributed on each sub-array along the $x\text{-}$ and the $y\text{-}$ axes, respectively. The frequency band is set from 26 - 30 GHz with 1500 frequency points. We consider three incident paths with parameters detailed in Table \ref{tab1}. The scan azimuth angle for MA is from $90^{\circ}$ to $270^{\circ}$, the scan elevation angle is from $0^{\circ}$ to $90^{\circ}$.

 		We first obtain beam pattern of the MA $B^0_{\scriptscriptstyle{MA}}(f,\theta,\phi)$ shown in Fig. \ref{BP}. It shows eight paths in the pattern. We expect nine paths corresponding to previous derivation that $K$ incident paths will result in $K^2$ detected paths. Among nine paths, two of them overlaps with each other, which will not influence the later detection. We detect the strongest path to obtain its elevation and azimuth angles which are $\{ \theta_1 =60^{\circ}, \phi_1=120^{\circ} \}$. Based on this angular position, we can estimate the delay by calculating the PADP at $\theta = \theta_1$, and reconstruct the CFR to estimate the power according to Algorithm 1. The CIR is gradually removed in each iteration shown in Fig. \ref{CIR}. After the first path been removed, the beam pattern $B^1_{\scriptscriptstyle{MA}}(f,\theta,\phi)$ shows only four paths, the fake paths introduced by first main path all vanishes. In the end, a path with power value within $30 \ \text{dB}$ dynamic range can not be found based on $\mathbf{h}^{2}_{\scriptscriptstyle{MA}}(\tau)$ and therefore the channel estimation procedure is complete. The numerical simulation results have demonstrated the effectiveness of the proposed algorithm.

		\begin{figure}[!t]
			\centering
			\includegraphics[width=0.47\textwidth]{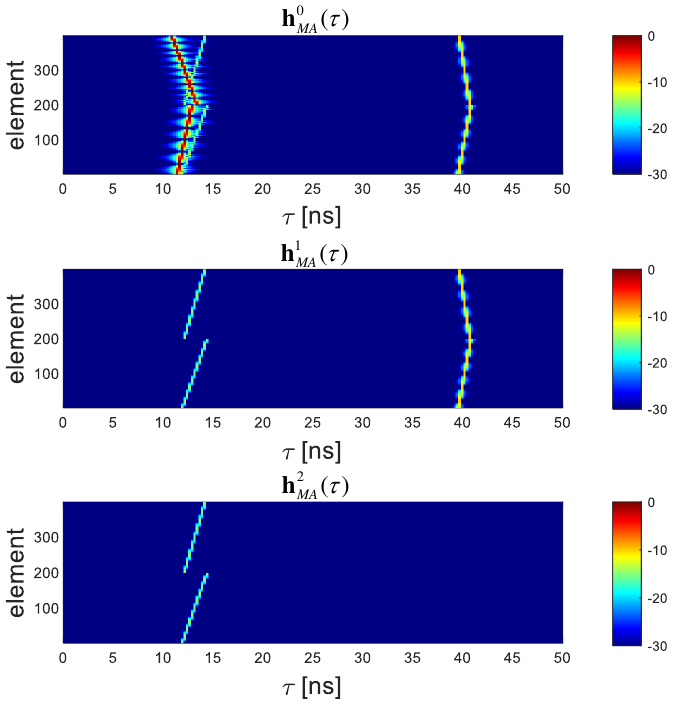}
			\caption{The CIR vectors of $\mathbf{h}^{q}_{\scriptscriptstyle{MA}}(\tau)$, $q \in [0,2]$, where the superscript denotes the $q$ path(s) are removed.}
			\label{CIR}
		\end{figure}
		
		\begin{table}[!t]
			\begin{center}
				\caption{Simulation path parameters} \label{tab1}
				\begin{tabular}{|c|c|c|c|c|}
					\hline 
					& Delay {[}ns{]} & \begin{tabular}[c]{@{}c@{}}Elevation\\ Angle [$^{\circ}$] \end{tabular} & \begin{tabular}[c]{@{}c@{}}Azimuth \\ Angle [$^{\circ}$] \end{tabular} & power {[}dB{]} \\ \hline
					path 1 & 12             & 60                                                        & 120                                                      & 0           \\ \hline
					path 2 & 40             & 30                                                        & 140                                                      & -10          \\ \hline
					path 3 & 13             & 80                                                        & 220                                                      & -15         \\ \hline
				\end{tabular}
			\end{center}
			
		\end{table}

		\section{Experimental Validation}
		\subsection{Measurement Campaign}
		
		\begin{figure}[!t]
			\centering
			\includegraphics[width=0.41\textwidth]{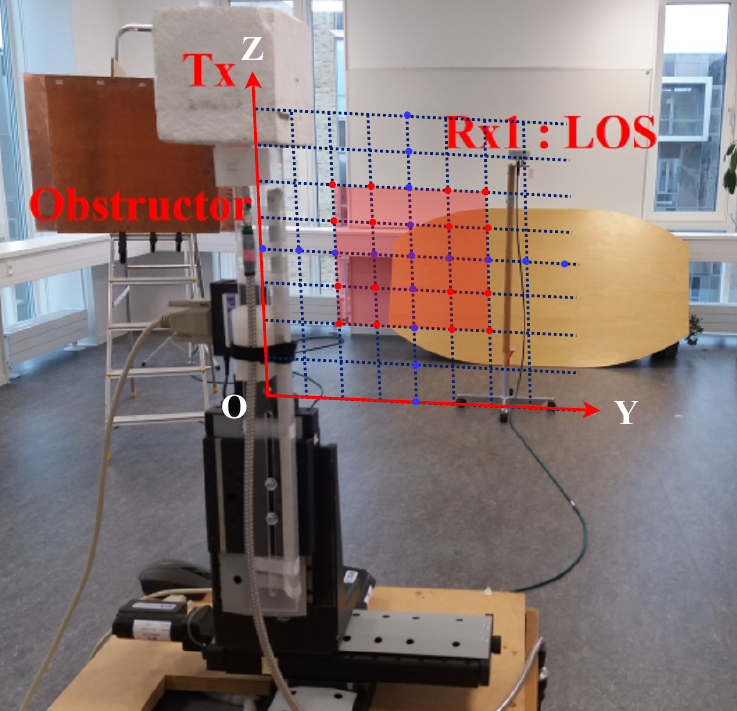}
			\caption{ Measurement scenario viewed from the Tx antenna.}
			\label{f4}
		\end{figure}
		
		To evaluate the practical efficacy of the proposed algorithm, validation measurements were carried out. A comprehensive description of the measurement campaign is provided in \cite{mbugua2018millimeter}, with only a brief overview presented here. The measurement was conducted in an empty 36 $m^2$ meeting room, and a static environment could be maintained for the entire duration of the measurement campaign. The measurement utilizes a center frequency of 27 GHz with 750 points across a 2 GHz bandwidth, yielding a delay resolution of 0.5 ns. The intermediate frequency (IF) bandwidth of vector network analyzer (VNA) was set to 500 Hz to enhance dynamic range, resulting in a sweep time of 2.89 s for each measurement location. Precision linear positioning stages are employed to construct VAA for measurement campaigns by moving the Tx antenna to predefined positions, as illustrated in Fig. \ref{f4}. The lengths of the linear positioning stages are 15 cm for the $x\text{-}$ and $y\text{-}$ axes, and 5 cm for the $z\text{-}$ axes \cite{mbugua2018millimeter}. The Tx and Rx antennas are both vertically polarized biconical antennas which have an omnidirectional pattern in the azimuth plane. 
		
		A back-to-back calibration procedure was conducted prior to undertaking the measurements. The virtual antenna array inter-element distance $d$ was set to 0.414$\lambda$ at 27 GHz to avoid grating lobes in the array pattern. Massive sampling is carried out and a total of (30 $\times$ 10) points for URA are recorded. We chose (5 $\times$ 15) and (9 + 29) points with the same phase center from these 300 points, representing the CFR for URA and MA, respectively. The recorded data for URA is processed with CBF algorithm \cite{ioannides2005uniform} and the data for the MA is processed with our proposed estimation algorithm.
		
		\begin{table*}[]
			\begin{center}
				\caption{Measurement of estimated parameters for URA and MA} \label{tab2}
				\begin{tabular}{|c|ccc|ccc|ccc|}
					\hline 
					\multicolumn{1}{|c|}{Parameters} & \multicolumn{3}{c|}{Delay [ns]}  & \multicolumn{3}{c|}{Azimuth Angle [$^\circ$]}   & \multicolumn{3}{c|}{Power [dB]}                                      \\ \hline 
					\diagbox{ \ \ No.}{Array}                                  & \multicolumn{1}{c|}{UPA}   & \multicolumn{1}{c|}{MA}    & Error & \multicolumn{1}{c|}{UPA} & \multicolumn{1}{c|}{MA}  & Error & \multicolumn{1}{c|}{UPA}    & \multicolumn{1}{c|}{MA}     & Error \\ \hline
					Path 1                                  & \multicolumn{1}{c|}{14.00} & \multicolumn{1}{c|}{14.25} & 0.25  & \multicolumn{1}{c|}{180} & \multicolumn{1}{c|}{180} & 0  & \multicolumn{1}{c|}{-65.1} & \multicolumn{1}{c|}{-65.8} & -0.7  \\ \hline
					Path 2                                  & \multicolumn{1}{c|}{21.00} & \multicolumn{1}{c|}{21.00} & 0.00  & \multicolumn{1}{c|}{180} & \multicolumn{1}{c|}{180} & 0  & \multicolumn{1}{c|}{-73.0} & \multicolumn{1}{c|}{-73.4} & 0.4  \\ \hline
					Path 3                                  & \multicolumn{1}{c|}{36.00} & \multicolumn{1}{c|}{36.25} & 0.25  & \multicolumn{1}{c|}{180} & \multicolumn{1}{c|}{181} & 1  & \multicolumn{1}{c|}{-74.7} & \multicolumn{1}{c|}{-73.6} & 1.1  \\ \hline
					Path 4                                  & \multicolumn{1}{c|}{19.50} & \multicolumn{1}{c|}{19.50} & 0.00  & \multicolumn{1}{c|}{142} & \multicolumn{1}{c|}{141} & -1  & \multicolumn{1}{c|}{-72.9} & \multicolumn{1}{c|}{-72.4} & 0.5  \\ \hline
					Path 5                                  & \multicolumn{1}{c|}{28.50} & \multicolumn{1}{c|}{28.75} & 0.25  & \multicolumn{1}{c|}{180} & \multicolumn{1}{c|}{180} & 0  & \multicolumn{1}{c|}{-78.4} & \multicolumn{1}{c|}{-75.9} & 2.5  \\ \hline
					Path 6                                  & \multicolumn{1}{c|}{25.50} & \multicolumn{1}{c|}{25.75} & 0.25  & \multicolumn{1}{c|}{229} & \multicolumn{1}{c|}{229} & 0  & \multicolumn{1}{c|}{-79.9} & \multicolumn{1}{c|}{-76.8} & 3.1  \\ \hline
				\end{tabular} 
			\end{center}
		\end{table*}

		\begin{figure*}[!t]
			\centering
			\includegraphics[width=1\textwidth]{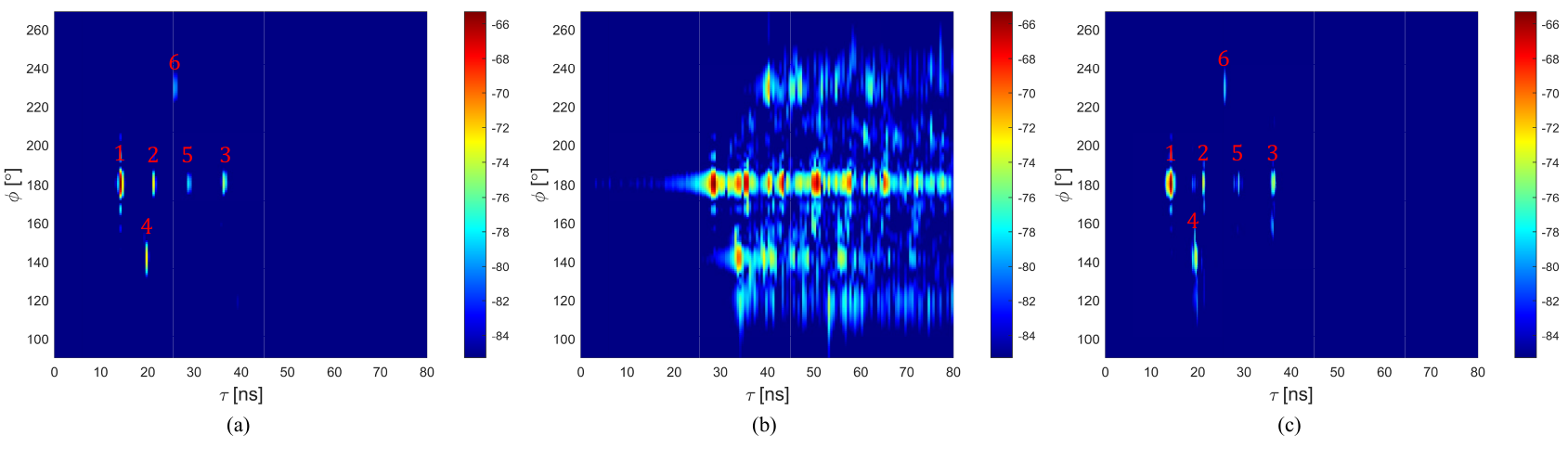}
			\caption{ (a) The PADP of the URA. (b) The PADP of the MA without using proposed algorithm. (c) The PADP of the MA with proposed algorithm.}
			\label{f5}
		\end{figure*}
		
		\begin{figure*}[!t]
			\centering
			\includegraphics[width=1\textwidth]{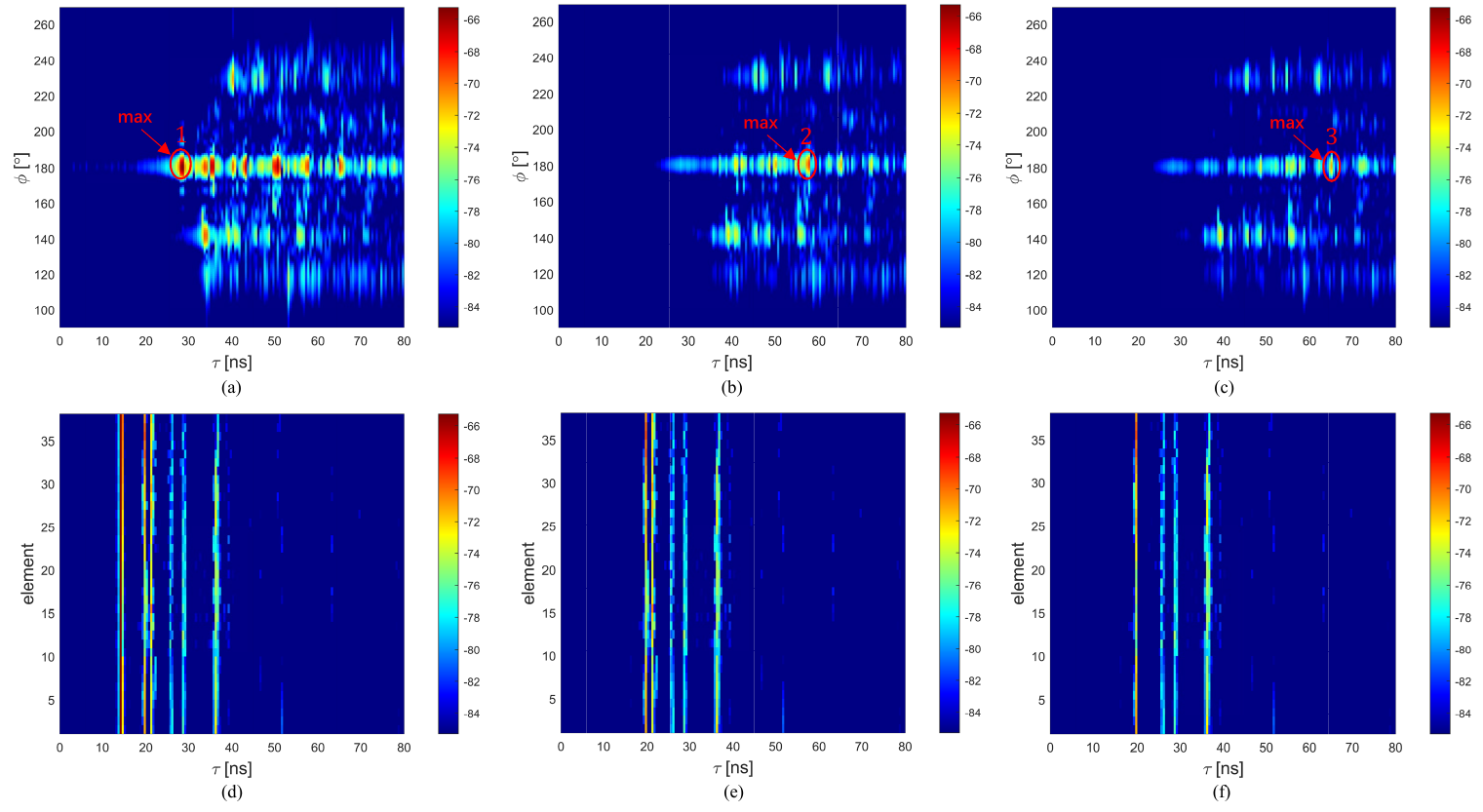}
			\caption{ The detailed PADP of the MA with corresponding CIR processed by proposed algorithm in first three iterations.  }
			\label{f6}
		\end{figure*}
		
		\subsection{Measurement results}

		Tx and Rx antennas are placed at a height approximately halfway between the floor and the ceiling, leading to large elevation angles in the propagation channel and elevation pattern of the deployed antennas is narrow. Therefore, multipaths that are not close to the azimuth plane are effectively filtered out. Therefore, the elevation angles are omitted from this analysis. 
		
		The spatial-temporal characteristic of the multi-path scenario in the LOS case is shown in Fig. \ref{f5}. Fig. \ref{f5} (a) demonstrate the results of URA with the CBF algorithm, it shown six distinct paths. In contrast, Fig. \ref{f5} (b) shows the results derived from a MA with the same CBF algorithm, wherein the spatial parameters appear indistinct and several spurious paths are observed. The presence of these erroneous paths significantly compromises the accuracy of channel parameter estimation. Upon applying our proposed algorithm to the MA, Fig. \ref{f5} (c) demonstrates that the results closely align with those of Fig. \ref{f5} (a), achieving an accurate estimation of channel parameters. The exact estimations of the primary LOS components for URA and MA are documented in Table \ref{tab2}, underscoring the high precision of our methodology. The angular and delay estimation accuracy for all paths are within 1$^{\circ}$ and 0.5 ns, respectively. We can observe that there only exists some deviation in path power values. As the number of iterations increases, the overall deviation exhibits a rising trend which may be caused by incomplete removal of detected path's CFR. The influence of residual paths on channel parameter estimation is exacerbated with each iteration, leading to progressively increasing bias in the power estimation. However, all dominant paths can be accurately estimated, which is sufficient to estimate dominant paths, especially at mmWave frequency bands.

		To further explain the process of our proposed algorithm, we present the detailed PADP of the MA with corresponding CIR in first three iterations after removing the detected path's CBF shown in Fig. \ref{f6}. We can find that the first main path with strongest power has large affect on other paths. Once it is removed, other fake paths introduced by the multiplication of this path with other true paths are all removed. The second strongest path emerges preparing for the next iteration. The SIC procedure will repeat until all dominant paths are extracted, as expected.

		\section{Conclusion}
		The MA has great potential to significantly reduce the number of required antennas for massive MIMO applications. This paper extends the application of the MA concept to channel sounding. We find that employing the MA for channel sounding will introduce $K^2-K$ fake paths with disordered angle and delay profiles when there exists $K$ incident paths. Based on this problem, we propose an channel parameter estimation algorithm based on SIC. By detecting strongest path, removing the reconstruction of single detected path's CFR in each iteration, we can realize high precision estimation of elevation angle, azimuth angle, delay and power of each multipath component. Both numerical simulation and experimental measurement has validated the effectiveness of the proposed algorithm. We have demonstrated that, with our proposed SIC algorithm for the MA, we can use a MA composed of two othogothal linear arrays of size $2M-1$ and $2N-1$ to replace a URA of $M \times N$ in channel sounding. This finding is significant, since we can employ the MA to dramatically reduce the system cost or improve the measurment efficiency, addressing one of the key bottlenecks in massive MIMO array based channel sounding. In the measurement campaign, we utilize 38 antenna elements in MA to realize the similar channel parameter estimation results as a URA consisting of 5 $\times$ 15 = 75 isotropic antenna elements, saving 37 elements (49.3 $\%$ thinning). The multipath parameters of all paths are estimated with high accuracy with the propsoed algorithm for the MA. Note that the traditional MA for array synthesis works reported in the literature can be viewed as an special case for our proposed MA framework under the assumption of narrowband and one single path. This work makes it possible to apply the MA concept for wideband multipath applications.

		\bibliographystyle{unsrt}
		\bibliography{reference_list}

		\newpage
		\vfill
		
	\end{document}